\documentclass[twocolumn,superscriptaddress,english]{revtex4}

\usepackage[T1]{fontenc}
\usepackage[utf8]{luainputenc}
\usepackage{amsmath,amssymb}
\usepackage{graphicx,subfig}
\usepackage{subscript}

\begin{document}

\title{Stationary patterns in star networks of bistable units: Theory and application to chemical reactions}
\author{Nikos E. Kouvaris} 
\affiliation{Center for Brain and Cognition, Universitat Pompeu Fabra, Barcelona, Spain.}
\affiliation{Department of Information and Communication Technologies, Universitat Pompeu Fabra, Ramon Trias Fargas, 25-27, 08005 Barcelona, Spain.}
\author{Michael Sebek}
\affiliation{Department of Chemistry, Saint Louis University, 3501 Laclede Ave., St. Louis, Missouri 63103, USA}
\author{Albert Iribarne} 
\affiliation{Departament de F\'isica de la Mat\`eria Condensada, Universitat de Barcelona, Mart\'i i Franqu\`es 1, 08028 Barcelona, Spain.}
\author{Albert D\'iaz-Guilera} 
\affiliation{Departament de F\'isica de la Mat\`eria Condensada, Universitat de Barcelona, Mart\'i i Franqu\`es 1, 08028 Barcelona, Spain.}
\affiliation{Universitat de Barcelona Institute of Complex Systems (UBICS), Universitat de Barcelona, Barcelona, Spain.}
\author{Istv\'an Z. Kiss} 
\affiliation{Department of Chemistry, Saint Louis University, 3501 Laclede Ave., St. Louis, Missouri 63103, USA}
\date{\today}

\begin{abstract}
We present theoretical and experimental studies on pattern formation with bistable dynamical units coupled in a star network configuration.
By applying a localized perturbation to the central or the peripheral elements, we demonstrate the subsequent spreading, pinning, or retraction of the activations; such analysis enables the characterization of the formation of stationary patterns of localized activity.
The results are interpreted with a theoretical analysis of a simplified bistable reaction-diffusion model. 
Weak coupling results in trivial pinned states where the activation cannot propagate. 
At strong coupling, uniform state is expected with active or inactive elements at small or large degree networks respectively. 
Nontrivial stationary spatial pattern, corresponding to an activation pinning, is predicted to occur at intermediate number of peripheral elements and at intermediate coupling strengths, where the central activation of the network is pinned, but the peripheral activation propagates toward the center.  
The results are confirmed in experiments with star networks of bistable electrochemical reactions. 
The experiments confirm the existence of the stationary spatial patterns and the dependence of coupling strength on the number of peripheral elements for transitions between pinned and retreating or spreading fronts in forced network configurations (where the central or periphery elements are forced to maintain their states).
 
\end{abstract}

\maketitle

\section{Introduction}
Complex networks can be formed by chemical reactors \cite{Horsthemke:PLA_2004, Karlsson:PNAS_2002, Toiya:ANGIE_2013, Kouvaris:ANGIE_2016, Kouvaris:PLOSONE_2012, Sebek:PRL_2016}, biological cells \cite{Bignone:JBiolPhys_2001, Varshney:PlosCompBiol_20011}, or engineered units \cite{Dorfler:2013:2005-2010, Ravoori:2011:034102, Newman:2003p6292}. 
The synergetic action of the dynamics in the nodes and the structure of the links results in a variety of self-organization phenomena including: Synchronization \cite{Arenas:PhysRep_2008, Baer:ANGIE_2012}, chimera states \cite{Tinsley:NatPhys_2012, Wickramasinghe:PlosONE_20013, Nkomo:PRL_2013, Hizanidis:SciRep_2016}, excitation waves \cite{Kouvaris:EPL_2014, Isele:NJP_2015, Steele:Chaos_2006}, stationary Turing \cite{Nakao:NatPhys_2010, Wolfrum:PhysD_2012, Tompkins:2014:4397-4402, Kouvaris:SciRep_2015} and oscillatory patterns \cite{Hata:SciRep_2014, Asllani:NatComm_2014}.  
Stationary patterns have also been found in networks of coupled bistable elements \cite{Kouvaris:PLOSONE_2012, Kouvaris:ANGIE_2016, Kouvaris:EPL_2013}.

Bistable behavior is encountered in many dynamical processes in chemical \cite{Mikhailov:Synergetics_I,  Epstein:NonlinearChemDyn_1998, Kouvaris:ANGIE_2016}, biological \cite{Graham:Develop_2010}, social \cite{Tess:PhysA_2005} and engineered systems \cite{Ikeda:PRL_1980}. 
In continuous bistable media, traveling fronts, representing waves of transition from one stable state into another can be observed \cite{Mikhailov:Synergetics_I, Epstein:NonlinearChemDyn_1998}. 
Traveling fronts can become pinned if coupling is sufficiently weak \cite{Booth:PhysA_1992, Erneux:PhysD_1993, Booth:JPhysChem_1994, Mitkov:PRL_1998, Laplante:PhysD_1992}, forming stationary patterns, in chains and lattices of diffusively coupled bistable elements.
Complex tree networks of coupled bistable units, both regular and irregular, exhibit the spreading, retreating or stationary patterns dependent on the coupling strength and the degree distribution of the nodes \cite{Kouvaris:PLOSONE_2012, Kouvaris:EPL_2013, Kouvaris:ANGIE_2016}. 

The current work was motivated by our previous study \cite{Kouvaris:ANGIE_2016}, where stationary 
pattern formation was observed in experiments with chemical bistable units based on a theory 
developed for tree (or tree-like) networks \cite{Kouvaris:PLOSONE_2012, Kouvaris:EPL_2013}. 
Intuitively, one could expect similar behavior in star network configuration. 
However, the previous theories \cite{Kouvaris:PLOSONE_2012, Kouvaris:EPL_2013} depend on a description of dynamics using three layers of units. 
In a star network, only two layers (center and periphery) are present and thus prediction of patterns with large number of elements in the nonlinear system provides a challenge. 

In this paper, we investigate the formation mechanisms of localized stationary patterns for star networks, where multiple bistable elements are connected to a central hub bistable unit.
This connectivity structure is often found in many natural or engineered systems that consist of dynamical elements interacting with each other through a common medium.
Examples include computer networks \cite{Roberts:1970} and optically coupled semiconductor lasers \cite{Zhang2008, Xiang2016, Zamora2010, Aviad2012, Bourmpos2012} where synchronization phenomena have been investigated. 
We present a theory that takes advantage of the simplicity of the star network topology to determine the conditions required for the formation of stationary patterns (without the approximations demanded for the bistable tree networks \cite{Kouvaris:PLOSONE_2012}).
Using the theory, we determined whether an initial activation will spread, retreat, or remain stationary for a given number of elements connected to the central unit and the strength of those connections. 
Finally, we designed an electrochemical star network system with bistable reaction units to confirm the theoretical findings and to demonstrate the existence 
of the network-topology induced stationary patterns.

\section{Star networks of coupled bistable units}
A simple model for a network organized bistable system can be given by the general form,
\begin{equation}\label{eq:brd}
\dot u_i = u(h-u)(u-a) + K\sum_{j=1}^{N}A_{ij}(u_j-u_i) \,,
\end{equation}
\noindent where $u_i$ denotes the amount of the activator in the $i$-th network node ($i=1,\ldots,N$), $0\!<\!h\!<\!a$ and the summation term represents the diffusive coupling between the nodes.
Parameter $K$ characterizes the coupling strength and $A_{ij}$ are the elements of the network's adjacency matrix with $A_{ij}=1$ if nodes $i$ and $j$ are connected and $A_{ij}=0$ otherwise. 
Therefore, the system \eqref{eq:brd} for the star networks can be formulated as,
\begin{subequations}\label{eq:sgsys}
\begin{eqnarray}
\dot u_1 & = & u_1(h-u_1)(u_1-a) + K\sum_{j=2}^{k+1}\!\left(u_j-u_1\right)\label{eq:sgsysa}\,,
\\
\dot u_i & = & u_i(h-u_i)(u_i-a) + K(u_1-u_i) \label{eq:sgsysb}\,, 2\leq i\! \leq\! k+1\,,
\end{eqnarray}
\end{subequations}
\noindent where Eqs. \eqref{eq:sgsysa} and \eqref{eq:sgsysb} describe the dynamics of the central and peripheral nodes respectively. 
Because of the symmetry in the system all peripheral nodes obey the same equation, thus the index $i$ can be dropped, and the system \eqref{eq:sgsys} is reduced into a two-dimensional system of ordinary differential equations,
\begin{subequations}\label{eq:redsys}
\begin{eqnarray}
\dot u & = & u(h-u)(u-a) + kK(v-u) := f(u,v;k,K)\,,\label{eq:redsysa}
\\
\dot v & = & v(h-v)(v-a) + K(u-v) := g(u,v;K)\,,\label{eq:redsysb}
\end{eqnarray}
\end{subequations}
\noindent where $u$ and $v$ denote the amount of activator in the central and peripheral nodes respectively.

In the absence of coupling ($K\!=\!0$), the dynamical system \eqref{eq:brd} has three fixed points: The stable nodes $u^*=0$ and $u^*=a$, and the saddle point $u^*=h$. 
In the following we will refer to the steady state $u^*=0$ as the {\it passive state} and to $u^*=a$ as the {\it active state} (c.f. \cite{Kouvaris:PLOSONE_2012, Kouvaris:ANGIE_2016}).

\section{Spatiotemporal dynamics driven by fixed boundary conditions}
We start our analysis by considering the system \eqref{eq:redsys} under two different fixed boundary conditions \cite{Erneux:PhysD_1993}. 
Firstly, the peripheral nodes are forced to be in the passive state.
Secondly, the central node is forced to be in the active state.
 
\subsection{Peripheral nodes forced to the passive state}
\begin{figure}[t!]
\includegraphics[width=\columnwidth]{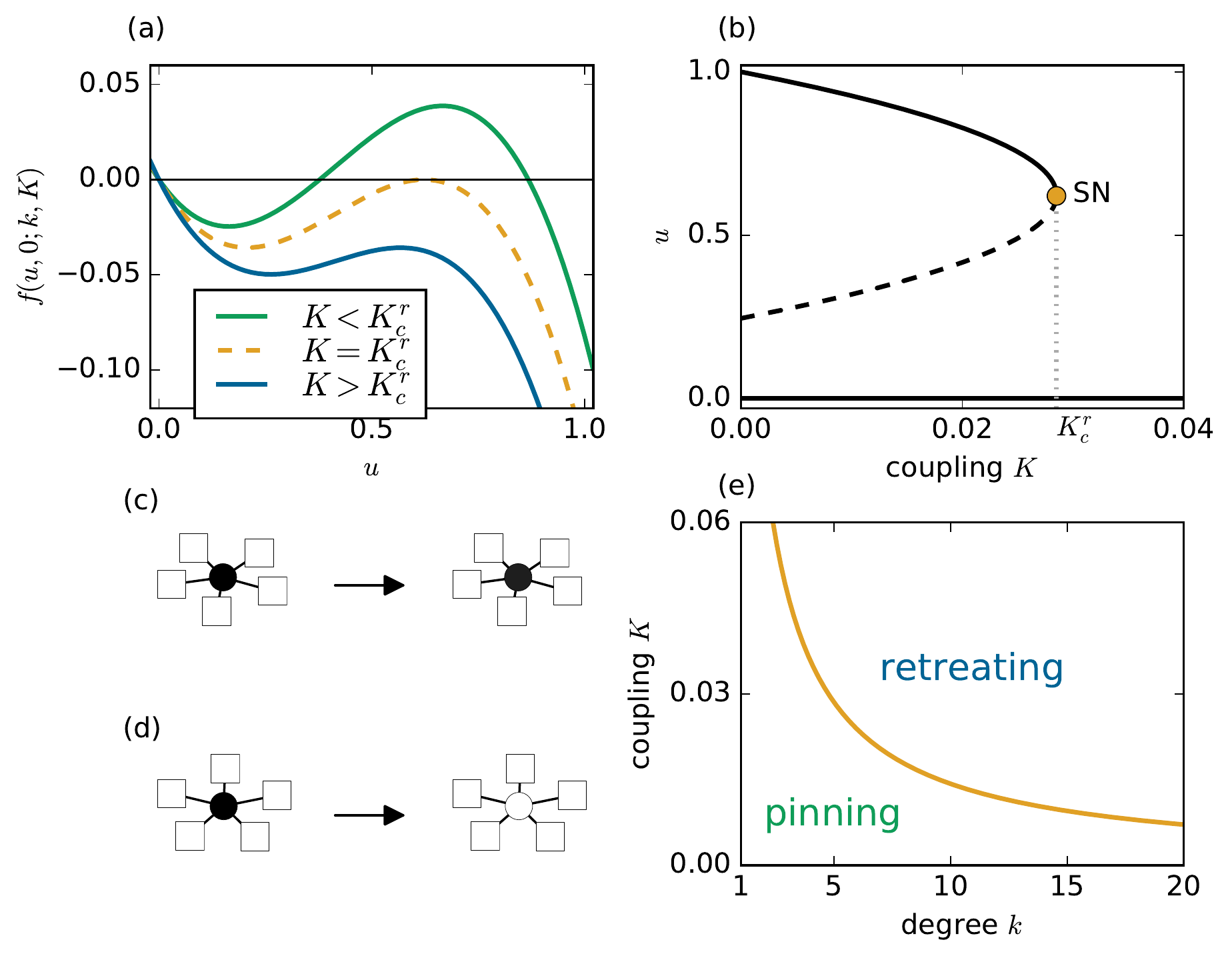}
\caption{Forced passive periphery: Bifurcation diagrams and simulations.
(a) The function $f(u,0;k,K)$ is plotted for $k=5$ and three different values of $K$. 
(b) The fixed points of Eq.~\eqref{eq:forceperiphery} are plotted as functions of $K$ for $k=5$.
(c) Initial activation remains stationary and localized in the center. 
(d) Initial activation retreats to the passive state.
Active and passive states are denoted by black and white respectively.
Circle denotes the bistable node and squares the forced nodes.
(e) The saddle-node bifurcation (thick curve) given by Eq.~\eqref{eq:snr} is shown in the $k$-$K$ parameters plane. Other parameters are $h=0.245$ and $a=1$.}
\label{triplefigurefr}
\end{figure}

We aim to determine the degree $k$ of the central node and the strength $K$ of the diffusive coupling which give rise to a stationary pattern representing localized activation of the central node while the peripheral nodes are forced to the passive state.
When the activation is initiated under the following initial and boundary conditions,
\begin{eqnarray}\label{eq:frini}
u & = & a \quad (t=0)\,,\nonumber\\
v & = & 0 \quad (t \geq 0)\,,\nonumber
\end{eqnarray}
\noindent the system \eqref{eq:redsys} is reduced into the ordinary differential equation, 
\begin{equation}\label{eq:forceperiphery}
\dot u = u(h-u)(u-a) - k K u \,,
\end{equation}
\noindent which describes the dynamics of the central node.
By solving $f(u,0;k,K)=0$ we find the fixed points of Eq.~\eqref{eq:forceperiphery},
\begin{eqnarray}
u_0 &:=& 0\,,\nonumber \\ 
u_{\pm} &:=& \frac{1}{2} \left(a+h\pm\sqrt{(a-h)^2-4 K k}\right)\,.\nonumber
\end{eqnarray}
\noindent Figure \ref{triplefigurefr}(a) shows that the fixed points $u_{+}$ and $u_{-}$ can vary with $k$ and $K$ and furthermore, they can merge and annihilate each other whereas $u_0$ always exists. 
Then the system has a critical coupling strength $K_c^r$,
\begin{equation}\label{eq:snr}
 K_c^r := \frac{(a-h)^2}{4 k}\,,
\end{equation}

\noindent for which a saddle-node bifurcation occurs  (Fig. \ref{triplefigurefr}(b)).
Analyzing the stability of the fixed points one can see that $u_0$ (solid line) is always stable, while $u_{+}$ (solid curve) and $u_{-}$ (dashed curve) are stable and unstable respectively.
If $K<K_c^r$ the trajectory starting at $u=a$ is attracted by the stable fixed point $u_+$, since $h<u_+<a$. 
This corresponds to a stationary activation localized on the central node (e.g. Fig. \ref{triplefigurefr}(c)).
After the saddle-node bifurcation, for $K>K_c^r$, every trajectory reaches the only stable fixed point $u_0$. 
This corresponds to the retraction of the initial activation (e.g. Fig. \ref{triplefigurefr}(d)).
In Figs. \ref{triplefigurefr}(c)--(d) circle denotes the bistable node and squares the forced nodes.
Figure \ref{triplefigurefr}(e) shows the saddle-node bifurcation in the $k$-$K$ parameters plane.

\subsection{Central node forced to the active state}
Here we look for stationary patterns when the central node is forced to be in the active state.
When the activation is initiated under the following boundary and initial conditions,
\begin{eqnarray}\label{eq:fpini}
u & = & a \quad (t\geq0)\,,\nonumber\\
v & = & 0 \quad (t = 0)\,, \nonumber
\end{eqnarray}
\noindent the system \eqref{eq:redsys} is reduced into the ordinary differential equation, 
\begin{equation}\label{fpg}
	\dot v = v(h-v)(v-a)+K(a-v)\,,
\end{equation}
\noindent which describes the dynamics of the periphery.
Following the same analysis as above we find the fixed points of Eq.~\eqref{fpg} by solving $g(1,v;K)=0$. This gives,
\begin{eqnarray}
v_a &:=& a\,, \nonumber\\ 
v_{\pm}&:=&\frac{1}{2}\left(h\pm\sqrt{h^2-4 K}\right)\,.\nonumber
\end{eqnarray}

\begin{figure}[t!]
\includegraphics[width=\columnwidth]{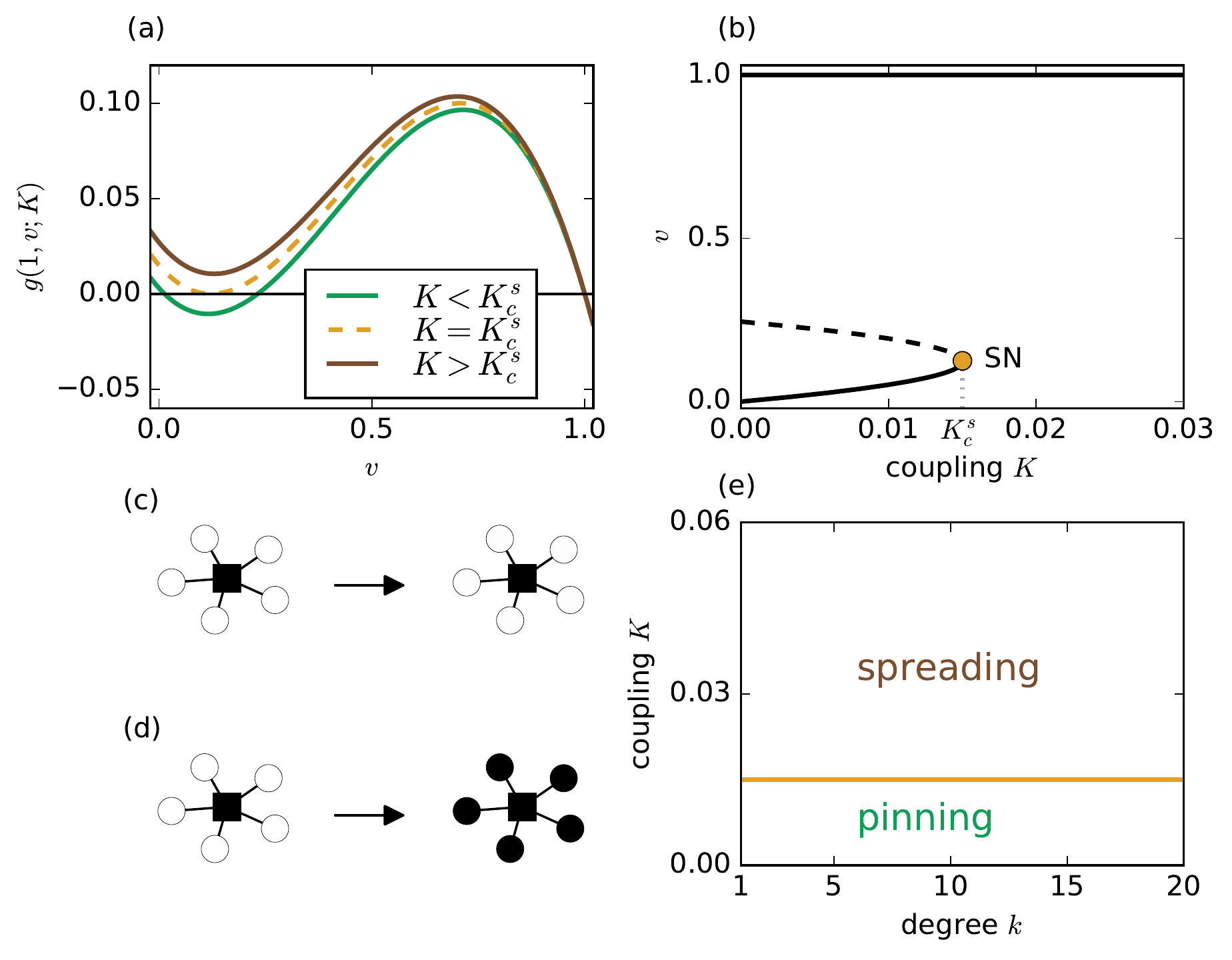}
\caption{Forced active center: Bifurcation diagrams and simulations.
(a) The function $g(1,v;K)$ is plotted for three different values of $K$. 
(b) The fixed points of Eq.~\eqref{eq:forceperiphery} are plotted as functions of $K$.
(c) Initial activation remains stationary and localized in the center. 
(d) Initial activation spreads to the periphery.
Active and passive states are denoted by black and white respectively.
(e) The saddle-node bifurcation (thick line) given by Eq.~\eqref{eq:snp} is shown in the $k$-$K$ parameters plane. Other parameters are $h=0.245$ and $a=1$.}
\label{triplefigurefp}
\end{figure}

Figure \ref{triplefigurefp}(a) shows that the fixed points $v_{+}$ and $v_{-}$ can vary with $K$ and, furthermore, they can merge and annihilate each other, whereas $v_a$ always exists. 
Then, we obtain the critical coupling strength $K_c^s$, 
\begin{equation}\label{eq:snp}
K_c^s := \frac{h^2}{4}\,,
\end{equation}
\noindent for which a saddle-node bifurcation occurs (Fig. \ref{triplefigurefp}(b)).
For every $k$, if $K<K_c^s$ there are three fixed points: The stable $v_a$ (solid line), the unstable $v_+$ (dashed curve) and the stable $v_-$ (solid curve). Hence the trajectory starting at $v=0$ is attracted by the stable fixed point $v_-$, since $0<v_-<h$. 
This corresponds to a stationary activation localized on the central node (e.g. Fig. \ref{triplefigurefp}(c)).
Otherwise, for $K>K_c^s$ there is only one (stable) fixed point at $v=a$. 
This correspond to the spreading of the initial activation to the peripheral nodes (e.g. Fig. \ref{triplefigurefp}(d)).
Figure \ref{triplefigurefp}(e) shows the saddle-node bifurcation in the $k$-$K$ parameters plane.

\section{Spatiotemporal dynamics triggered by localized initial perturbations}
The above analysis reveals that under fixed boundary conditions, an initial activation can be stationary and localized in the central node, can spread to the peripheral nodes bringing the whole network to a fully activated state, or can retreat prompting a passive state.
Now we analyze the general case, boundary conditions not fixed, where any of those steady states can be reached depending on the combination of the coupling strength $K$, the degree $k$ and the initial conditions.

The system~\eqref{eq:redsys} has three trivial fixed points at $u=v=0$, $u=v=a$ and $u=v=h$. 
The points $(0,0)$ and $(a,a)$ are stable nodes for any (positive) value of $k$ and $K$. 
They represent a star network with all nodes in the passive or in the active state respectively.
The point $(h,h)$ is an unstable node for $K<(a-h)h/(k+1)$ and becomes a saddle point for $K>(a-h)h/(k+1)$. 

The system~\eqref{eq:redsys} also has other fixed points, different from the three trivial ones mentioned before.
By solving $f(u,v;k,K)=0$ we can get an expression of $v$ in terms of $u$.
By substituting this expression into the equation $g(u,v;K)=0$, we obtain the 6th degree polynomial equation,
\begin{equation}\label{eq:poly}
P(u) := c_0 + c_1 u + c_2 u^2 + c_3 u^3 + c_4 u^4 + c_5 u^5 + u^6 = 0\,,
\end{equation}
\noindent whose coefficients in terms of $h$, $a$, $k$ and $K$ are shown in Table \ref{tab:coeff}.

\begin{table}[b!]
\begin{ruledtabular}
\begin{tabular}{ll}
$c_0 = K^2 k^2 (K+ a h + K k)$&\\
$c_1 = -K k (a + h) (a h + 2 K k)$& \\
$c_2 = 5 a K k h + a^2 (h^2 + K k) + K k (h^2 + 3 K k)$ &\\
$c_3 = -2 (a + h) (a h + 2 K k)$&\\
$c_4 = a^2 + 4 a h + h^2 + 3 K k$ &\\
$c_5 =-2 (a + h)$ &\\
\end{tabular}
\end{ruledtabular}
\caption{\label{tab:coeff}The coefficients of the polynomial $P(u)$}
\end{table}

Assuming that $u_*$ is a real root of $P(u)$, then $(u_*,v_*)$ is a fixed point of the system \eqref{eq:redsys}, where $v_*$ satisfies the equation $f(u_*,v_*;k,K)=0$.
When $(u_*,v_*)$ is stable, it corresponds to a steady state that attracts any initial activation found in its basin of attraction. 
Then, appropriate initial conditions can give rise to stationary patterns localized in the center or the periphery.
However, such a stable point can merge with a saddle point and disappear through a saddle-node bifurcation resulting in a transition from those localized patterns to activation spreading or retreating.
Unlike the cases of fixed boundary conditions the expression of this bifurcation cannot be obtained analytically; however, it can be numerically  determined from the conditions,
\begin{equation}\label{eq:sdgencond}
P(u)=0 \quad \text{and}\quad \frac{d}{du}P(u)=0\,.
\end{equation} 

\begin{figure}[t!]
\includegraphics[width=\columnwidth]{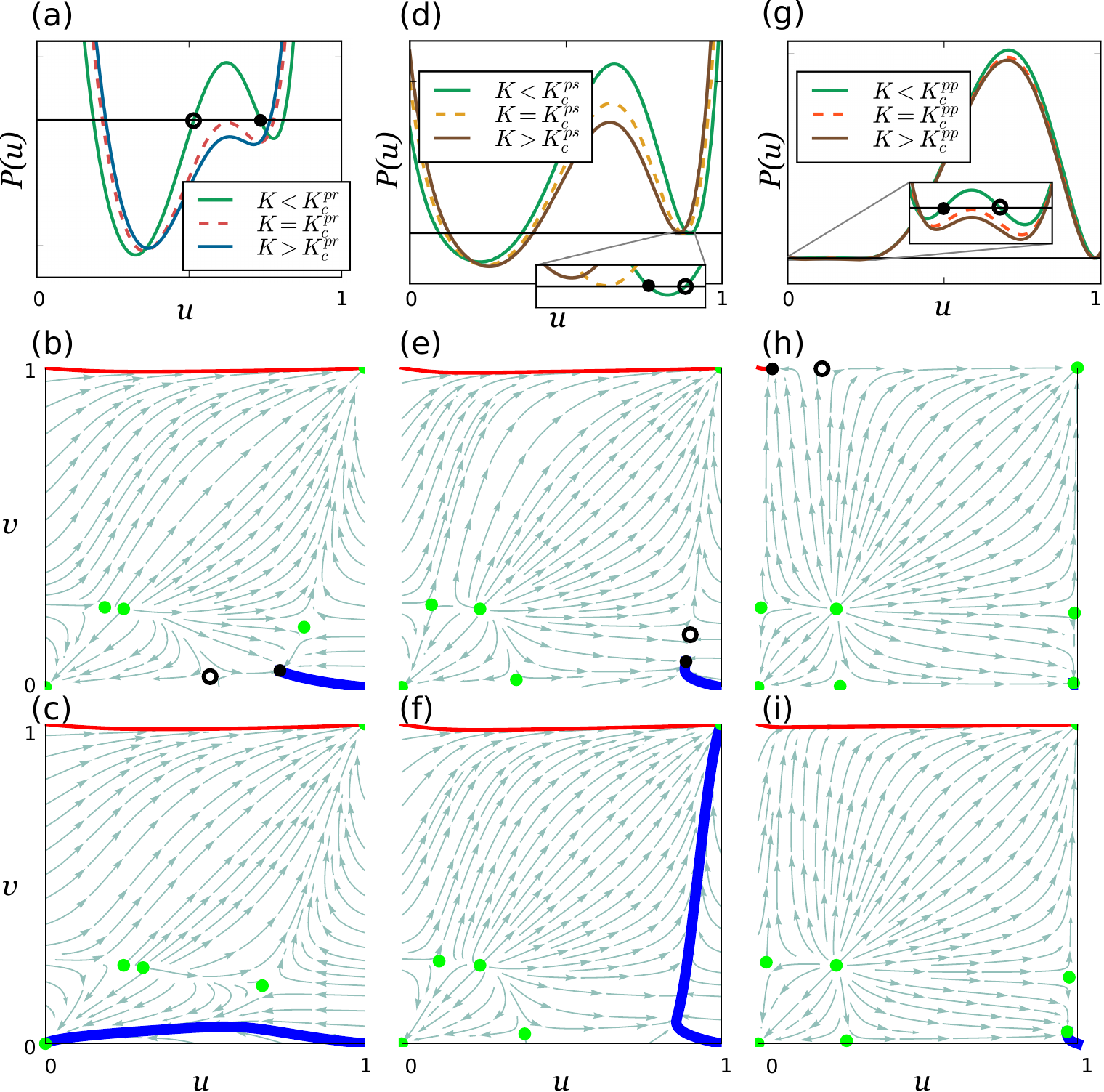}
\caption{Phase portraits before and after the saddle-node bifurcations.
(a) The polynomial $P(u)$ is plotted for $k=10$ and three different values of $K$.
The phase portrait and the vector field of system \eqref{eq:redsys} are shown for $k=10$ and (b) $K<K_c^{pr}$ and (c) $K>K_c^{pr}$.
The dots denote the fixed points, the thick trajectory represents the evolution of the initial center activation $(a,0)$, and the thin trajectory represents the evolution of the initial periphery activation $(0,a)$. 
Similar plots are presented for (d)--(f) $k=5$, and (g)--(i) $k=3$. Other parameters are $h=0.245$ and $a=1$.}
\label{fig:polyplot}
\end{figure}
  
Figure \ref{fig:polyplot}(a) shows $P(u)$ for $k=10$ and three values of coupling strength $K$. 
At $K_c^{pr}$ (dashed curve) the two inner roots (shown with circle and dot) of $P(u)$ merge together and disappear through a saddle-node bifurcation.
These roots correspond to the fixed points of Eqs.~\eqref{eq:redsys}, which are depicted with the same symbols in Fig.~\ref{fig:polyplot}(b).
Any initial condition $(u_0,v_0)$ in the basin of attraction of a stable fixed point will eventually 
converge to the corresponding fixed point. 
Therefore, before the bifurcation, the initial center activation $(a,0)$ results in a stationary pattern (thick trajectory in Fig.~\ref{fig:polyplot}(b)), and the initial periphery activation $(0,a)$ propagates towards the center (thin trajectory in Fig.~\ref{fig:polyplot}(b)).
After the bifurcation, the center activation retreats to the passive state (thick trajectory in Fig.~\ref{fig:polyplot}(c)) whereas the periphery activation propagates towards the center (thin trajectory in Fig.~\ref{fig:polyplot}(c)).
Figures \ref{fig:polyplot}(d)--(f) and (g)--(i) illustrate similar scenarios for $k=5$ and $k=3$ respectively.

\begin{figure}[t!]
\includegraphics[width=0.9\columnwidth]{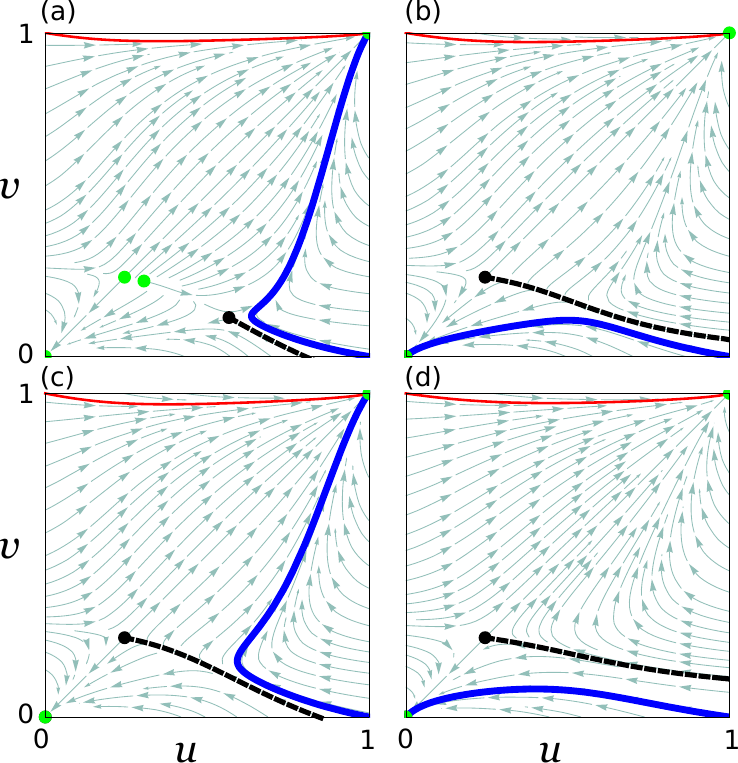}
\caption{Phase portraits beyond the cusp point.
The phase portrait and the vector field of system \eqref{eq:redsys} for $h=0.245$ and $a=1$. 
The dots denote the fixed points, the thick trajectory represents the evolution of the initial center activation $(a,0)$, and the thin trajectory represents the evolution of the initial periphery activation $(0,a)$. Black dashed trajectory shows the stable invariant manifold of the nearest saddle point to $(a,0)$ for
(a) $k=6$ and $K=0.0295$, 
(b) $k=6$ and $K=0.034$, 
(c) $k=5$ and $K=0.04$, and
(d) $k=7$ and $K=0.04$.}
\label{fig:phasespaceinv}
\end{figure}

Beyond the saddle-node bifurcations described in Fig.~\ref{fig:polyplot}, the invariant manifolds of the remaining saddle points change positions by varying $k$ or $K$.
Therefore, the partition of the phase space into different basins of attraction also changes giving rise either to a fully active or to a fully passive state, depending on the initial conditions.
Figures \ref{fig:phasespaceinv}(a)--(b) illustrate this behavior for varying $K$ where the relative position of one stable manifold (dashed trajectory) and the trajectories resulting from the initial center (thick trajectory) or periphery (thin trajectory) activation are shown.
Figures \ref{fig:phasespaceinv}(c)--(d) present the same scenario for varying $k$.

\begin{figure}[t!]
\includegraphics[width=\columnwidth]{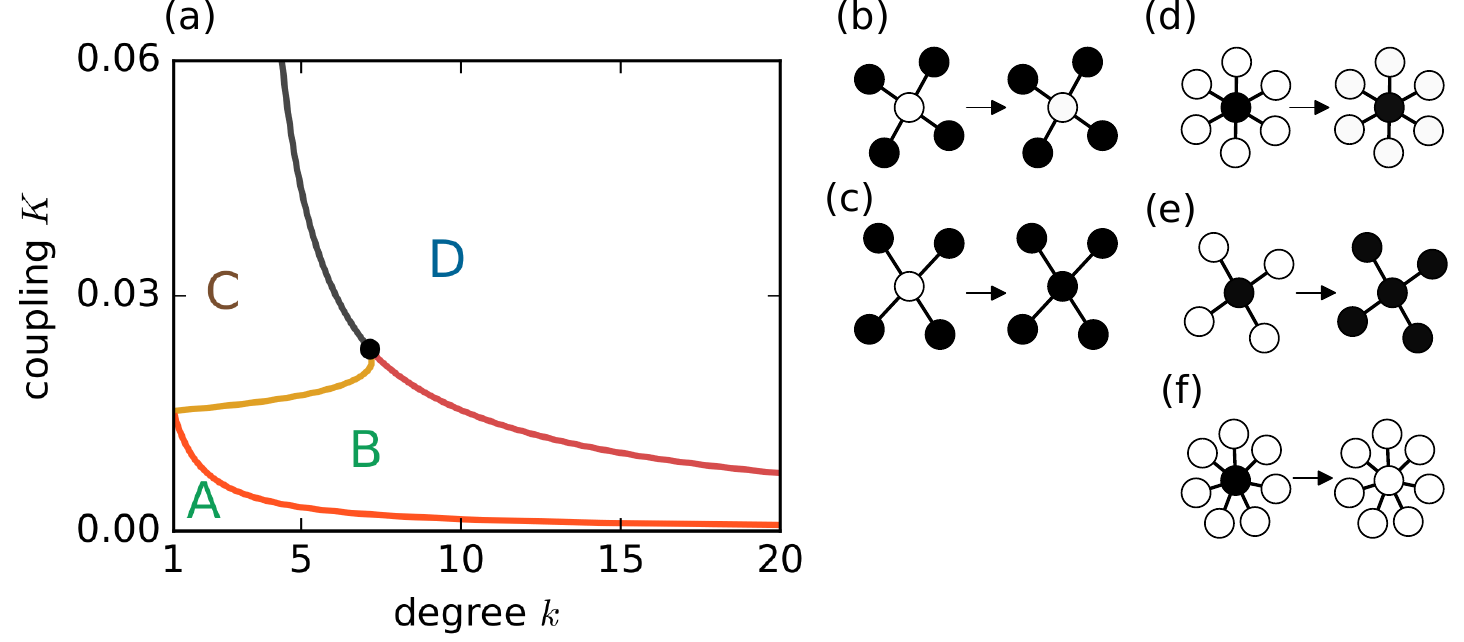}
\caption{Phase diagram and regional dynamics.
(a) Four different regions are shown in the $k$-$K$ parameters plane for $h=0.245$ and $a=1$.
Region A: Activations are pinned.
Region B: Center activation is pinned, and periphery activation propagates towards the center.
Region C: Center and periphery activations propagate. 
Region D: Center activation retreats, and periphery activation propagates.
The spatiotemporal evolution of (b)--(c) periphery and (d)--(f) center activations corresponding to different regions is also presented.
Simulations were performed for the parameters
(b) $k=4,\ K=0.001$, 
(c) $k=4,\ K=0.02$, 
(d) $k=6,\ K=0.005$, 
(e) $k=4,\ K=0.01$, 
(f) $k=7,\ K=0.04$.
Active and passive states are denoted by black and white respectively.}
\label{fig:regions}
\end{figure}

The aforementioned saddle-node bifurcations determine the partitions between stationary patterns and activation spreading, or between stationary patterns and activation retreating.
The stable invariant manifold of the nearest saddle point to $(a,0)$ (dashed line in Fig.~\ref{fig:phasespaceinv}) separates activation spreading from retreating.
Figure \ref{fig:regions}(a) summarizes this scenario in the $k$-$K$ parameters plane.
The curve separating regions B and D corresponds to the bifurcation described in the Figs.~\ref{fig:polyplot}(a)--(c); it represents the continuation of $K_c^{pr}$ for varying $k$. 
Similarly, regions B and C are separated by the bifurcation (at $K_c^{ps}$) shown in the Figs.~\ref{fig:polyplot}(d)--(f) whereas regions A and B by the bifurcation (at $K_c^{pp}$) shown in Figs.~\ref{fig:polyplot}(g)--(i).
We also see that the saddle-node bifurcations which enclose region B merge in the cusp point (thick point) which can be defined by the conditions $P(u)=(d/du)P(u)=(d^2/du^2)P(u)=0$.
The curve separating the areas C and D after the cusp bifurcation corresponds to the stable manifold of the saddle point that passes from $(a,0)$ as discussed in Fig. \ref{fig:phasespaceinv}.
Our analysis has also revealed that this curve tends asymptotically to the value $k=3$. 
This means that the center activation cannot retreat in a star network with $k\leq3$. 
We note that the general shape of the bifurcation diagram is very similar to those obtained for a tree networks \cite{Kouvaris:PLOSONE_2012}, i.e., the diagram has the same domains and the domains have the same relative position to each other. 
However, the exact position of the transition boundaries are different. 
For example, tree networks with $k=3$ exhibited retreating activation, which is not possible with the star network. 
Similarly, at the same nonlinearity parameter, the strongest coupling strength at which pinning states can exist (i.e., at the cusp point) $K=0.071$ in contrast with $K=0.023$ for the star network.

\section{Experiments with coupled bistable electrochemical reactions}
We designed an experimental setup with star networks of coupled bistable electrochemical reactions to confirm the aforementioned theoretical findings.
Each node in the network is represented by a nickel wire. 
At sufficiently large potential, the nickel wire undergoes transpassive dissolution \cite{haimmodel}. 
The rate of metal dissolution (or the corresponding potential drop across the electrical double layer driving the reaction) can be used as an experimental variable to characterize the state the of system. 
During the reaction, the metal ions dissolve through the passivating oxide layer on the surface; the rate of the dissolution can be inhibited by adsorbed bisulfate ions and through change in the composition of the oxide species on the surface \cite{haimmodel}.
These processes occur at sufficiently large potentials, and thus can slow down the dissolution and create a negative differential resistance (NDR) in the current vs. potential diagram.  
When sufficiently large series resistance is present in the system (e.g., through attached external resistance), the system can exhibit bistability between high current (low  electrode potential) and low current (high electrode potential) states at a given circuit potential. 
The electrodes can be coupled by attaching cross resistance between the peripheral and the (arbitrarily selected) central node \cite {Wickramasinghe:PlosONE_20013}. 
When the electrodes are coupled, current can flow between them in the presence of an electrode potential difference; this cross-current induces coupling by affecting the rate of metal dissolutions of the coupled electrodes.

\subsection{Experimental Setup}
\begin{figure}[t!]
\includegraphics[width=\columnwidth]{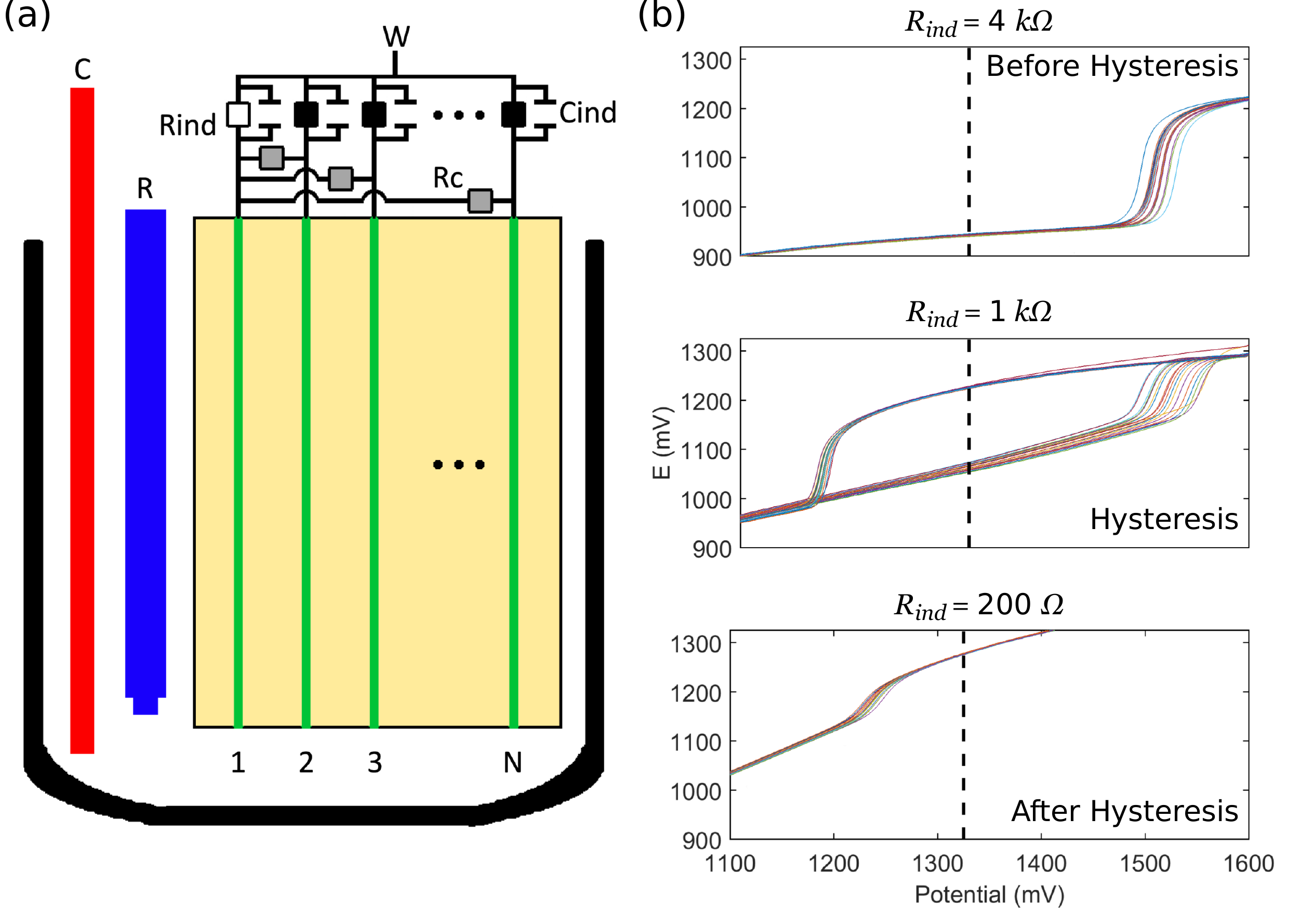}
\caption{Experimental setup and hysteresis of forced and bistable elements.   
(a) Experimental apparatus: $C$ is the counter electrode, $R$ is the reference electrode and $W$ is the working electrode array. 
An individual resistance ($R_\text{ind}$) and capacitance ($C_\text{ind}$) were connected in parallel to each electrode in the array. 
(b) The constant applied potential of experiments is before the hysteresis for forced active ($R_\text{ind}=4$ kOhm) and is after the hysteresis for forced passive ($R_\text{ind}=200$ Ohm) elements. 
Bistable elements occur when the constant potential ($V=1325$ mV) is within the hysteresis region ($R_\text{ind}=1$ kOhm).}
\label{fig:exp_setup}
\end{figure}

The experiments were performed using an electrochemical cell with a platinum coated titanium rod as the counter electrode, Hg/Hg\textsubscript{2}SO\textsubscript{4}/saturated K\textsubscript{2}SO\textsubscript{4} as the reference electrode, an array of twenty five  1.00\ mm diameter nickel wires as the working electrode. The cell electrolyte was 3 M H\textsubscript{2}SO\textsubscript{4} held at 10\textsuperscript{o}C (Fig.~\ref{fig:exp_setup}(a)). 
The electrodes in the array are connected to the potentiostat through an individual resistance ($R_\text{ind}$) and individual capacitance ($C_\text{ind}$) in parallel. 
The individual resistance provides the sufficient 
ohmic drop for bistability; the $C_\text{ind}$ serves to prevent the oscillations which can occur due to cell instabilities. 
The current of each electrode ($i_k$) is measured at 50 Hz data acquisition rate at a constant circuit potential $V$ 
(all potentials are given with respect to the reference electrode). 
The electrode potential ($E_k$) for each electrode is calculated by subtracting the ohmic potential drop on the individual resistance from the applied circuit potential,  
\begin{equation}\label{eq:itoE}
E_k :=  V- R_\text{ind} i_k \,.
\end{equation}

\noindent When $R_\text{ind}=1$ kOhm external resistance is attached to the wires, bistability occurs in a potential range of about 1200 mV $<V<$ 1500 mV (Fig.~\ref{fig:exp_setup}(b)). 
The experiments were performed at about $V\!=\!1300$ mV. 
Because the bistability range changes somewhat during the experiments, before each set of recorded data we determined the bistability range and the set the circuit potential to the middle of the lower half of the range.
During the experiments, the desired initial conditions (active or passive state of each electrode) were set by superimposing a locally applied potential sweep on the circuit potential. 

When the individual resistance was set to lower value of $R_\text{ind}=200$ Ohm, at the same circuit potential that corresponds to the bistable regime ($V\approx 1300$ mV), the electrodes are all passivated and exhibit the low current, high electrode potential state (see Fig.~\ref{fig:exp_setup}(b)). 
Because now the circuit potential is above the bistability (hysteresis) region, the system exhibits monostable behavior; we refer to this state as a ``forced'' passive state. 
Similarly, when the individual resistance was set to a higher value of $R_\text{ind}=4$ kOhm, at the same potential, the system is below the bistability region, and the electrodes exhibit a high current, low electrode potential state as shown in Fig.~\ref{fig:exp_setup}(b). 
We refer to this state as a ``forced'' active state. 
Because the passivation of nickel occurs at an electrode potential of about 1150 mV, we will use this threshold for classifying the state of the system into active ($E_k \!<\! 1150$ mV) or passive ($E_k \!>\! 1150$ mV). 
We note that the forced active and passive states are approximations of a theoretical forced active and passive states where the state of the system is not changed by any perturbation. 
In the experiments (as it will be shown below) there will be small changes of the forced states due to coupling, relative to the large changes we can observe with bistable units. 

Star network topologies are applied between electrodes via charge flow through connections of external coupling resistances ($R_{c}$) as shown in Fig.~\ref{fig:exp_setup}(a). 
The strength of the interactions ($K$) are given as the inverse of the the coupling resistance, i.e., $K=1/R_c$.
 
Whether the node is bistable (circle) or forced (square) to a particular state is dependent on the individual resistance.
In the following network diagrams the nodes represent the electrodes and the links the external connections between them.

\subsection{Experimental Results}
\begin{figure}[t!]
\includegraphics[width=\columnwidth]{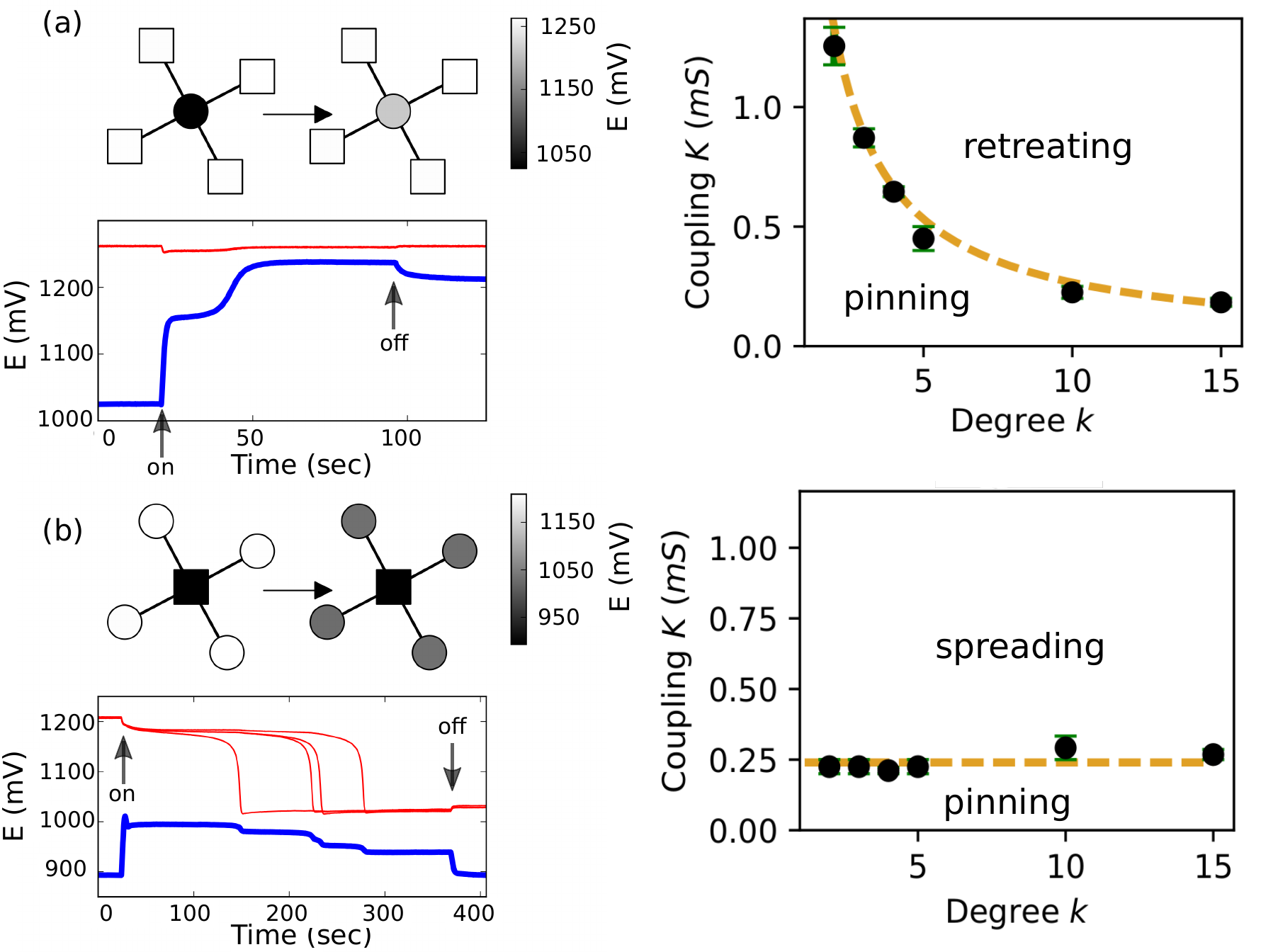}
\caption{Spatiotemporal dynamics of center activation in star networks with bistable (circle) and forced (square) elements. 
Initial and final states, the corresponding electrode potential time series and the resulting phase diagrams $k$-$K$ are shown for 
(a) forced passive periphery ($K=0.667$ mS) and
(b) forced active center ($K=0.222$ mS).
In the left panels the thick lines correspond to the central node and thin lines to the peripheral nodes.
The arrows indicate where the coupling was applied (on) and removed (off). 
In the right panels the dashed lines are fit to the theoretically predicted equations.
Experiments were performed at $V=1325$ mV.}
\label{fig:expforced}
\end{figure}

First, experiments with star networks with forced central or peripheral elements were undertaken.
The left panel in Fig. \ref{fig:expforced}(a) shows that a center activation retreats via the coupling to the forced passive periphery.
After turning the coupling on, we see large immediate increase followed by a slow relaxation of the electrode potential of the central unit to high potential (passive) state. 
Such transition can be observed only for sufficiently strong coupling; at weak coupling the activations are pinned. 
We determined the critical coupling strength needed to achieve the transition between the pinned and retreating activations. 
The right panel in Fig. \ref{fig:expforced}(a) shows that as the degree of the central node was increased, the critical coupling strength needed to achieve retreating fronts decreases. 
In this panel the points show the experimental results whereas the dashed curve is a least square fit to the theoretical prediction (Eq. \eqref{eq:snr}) that the coupling strength is proportional to the inverse degree, i.e., $K_c^r \propto 1/k$ with a value for $a-h=3.26$ mS.  

Similar experiments were performed with a forced active center and passive bistable periphery elements.
As shown in the left panels of Fig. \ref{fig:expforced}(b), sufficiently strong coupling results in spreading of the activation from the periphery. 
After the initial activation, it takes about 150 to 300 seconds for the periphery elements to reach the active states (small heterogeneity in the size of the bistable regions could contribute to the different transition times of the different electrodes).
The right panel of Fig. \ref{fig:expforced}(b) shows that, as predicted by the theory, the critical coupling strength for these activations do not depend on the degree and require a mean value of $K_c^s = 0.24 \pm 0.03 $ mS. 
Therefore, using Eq. \eqref{eq:snp} we found that $h=0.98$ mS.
By combining the $a-h=3.26$ mS value (form the forced passive periphery experiments) with $h=0.98$ mS  (from the forced active center experiments) the relative distance from the active state to the saddle point can be estimated to $h/a=0.23$.

\begin{figure}[t!]
\includegraphics[width=\columnwidth]{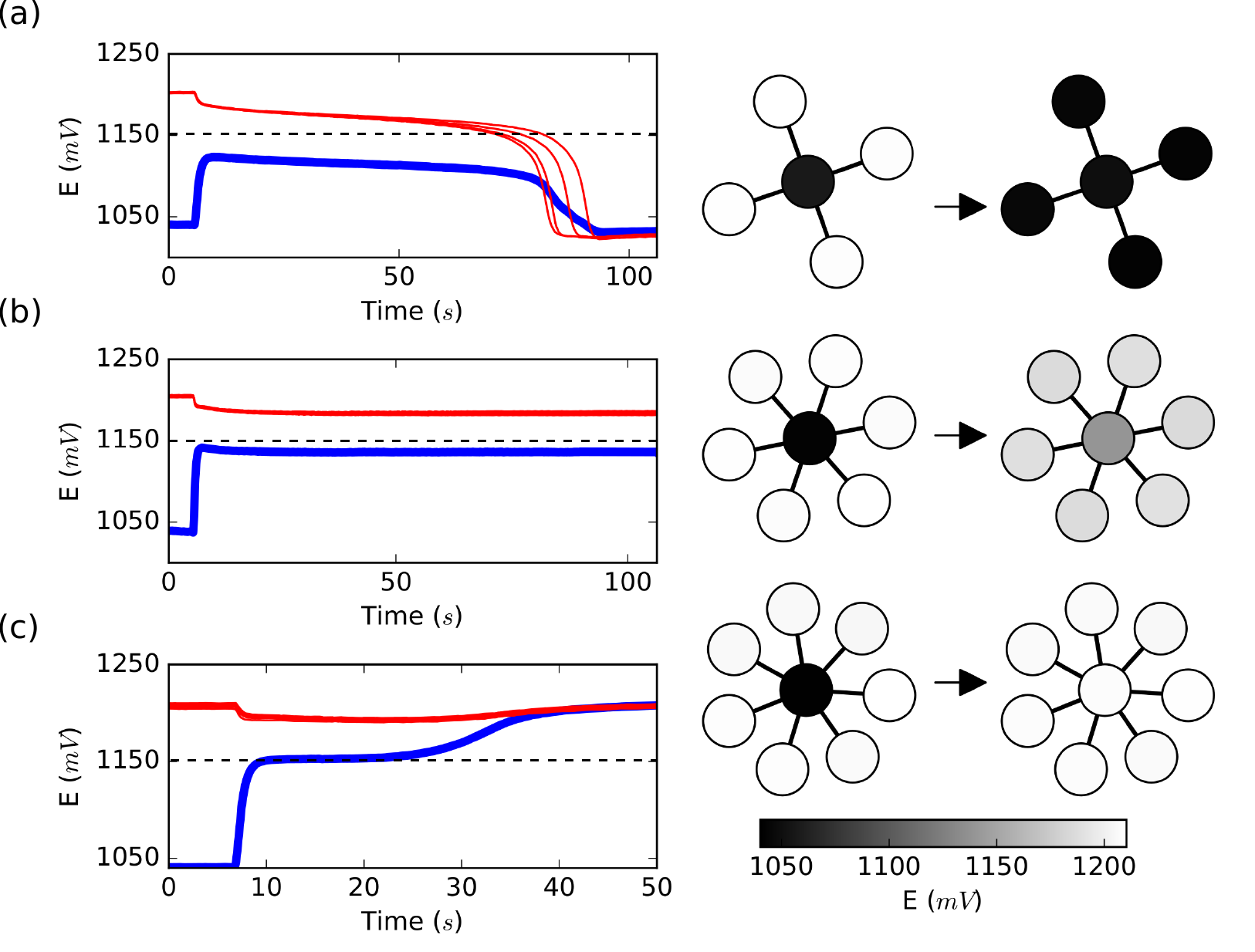}
\caption{Spatiotemporal dynamics of center activation in star networks of bistable elements. 
Electrode potential time series and the corresponding initial and final states of the star networks are shown for
(a) a spreading activation where the periphery (thin lines) activate via coupling to the center (thick line),
(b) a stationary activation between the peripheral and central elements,
(c) a retreating activation where the center passivates via coupling to the periphery. Experiments were performed at $V=1301$ mV and $K=0.667$ mS}
\label{fig:expgen}
\end{figure}

Finally, experiments where all elements, central and peripheral, are bistable were also performed to identify parameter region where non-trivial stationary pattern could arise. 
The coupling strength was set to value ($K=0.667$ mS) larger than that required for the ``trivial'' pinning state in the forced active center experiments ($K=0.24 \pm 0.03$ mS).
At this coupling strength, the experiments with forced passive periphery elements predicted that a four degree star network would be close to the transition between the pinning and retreating regions. 
The initial activation of the star network with $k=4$ shown in Fig. \ref{fig:expgen}(a) spreads and activates the peripheral nodes.
Therefore, the general shape of the phase diagram is correct, the pinning state is expected with higher degree network. 
By increasing the degree to $k=6$ and using the same coupling strength this center activation becomes pinned (Fig. \ref{fig:expgen}(b)) and thus a stationary spatial pattern was observed. 
For an even larger degree, $k=7$, the center activation retreats to the passive state (Fig. \ref{fig:expgen}(c)), as predicted by the theory.

The time series in Figs. \ref{fig:expgen}(a)--(c) also allow the estimation the relative position of the two stable and the one unstable steady states for the uncoupled systems, that correspond to values $0$, $a$, and $h$ respectively. 
The two stable states are located at about $E_a=1050$ mV (active) and $E_p=1200$ mV (passive) respectively. 
As it is shown in the Figs. \ref{fig:expgen}(a)--(c) with dashed lines, $E_h=1150$ mV is a good approximation of the saddle point as this value is close to the pinned states. 
Therefore, the distance between the passive and the saddle states relative to the distance between the passive and the active states can be obtained as $|E_h - E_p|/|E_a - E_p| = 0.33 $.
Note that this compares very well with the value obtained from the forced activation experiments (0.23). 
In addition, because the potential was set in the middle of the lower half of the bistability region, we can also approximate value with linear, Z-shaped hysteresis, which would give a value of 0.25.
In summary, the experimental findings are consistent with the theoretical predictions with a value of nonlinearity parameter $h \approx 0.2-0.3$.

 \section{Conclusions}

We have theoretically and experimentally demonstrated that bistable star networks support stationary patterns and activation spreading or retreating determined by: The number of coupled elements to the central unit, the coupling strength and the initial conditions. 
The theory demonstrates that stationary localized patterns are formed by the pinning of an initial activation.
The qualitative features of the phase diagram corresponding to the pinned, spreading and retreating activations, are the same as those observed for the tree networks although the three-layer approximation used for the trees does not directly apply for the star configurations \cite{Kouvaris:PLOSONE_2012}. 
While the qualitative features of the phase diagram are the same (number of domains and their relative placement), there are important quantitative difference between the tree and the star configurations. 
For example, at the same nonlinearity parameter, at strong coupling retreating fronts require larger degree networks for star topology than for a tree. 
Similarly, stationary patterns formed by pinned activation states survive stronger coupling with tree networks than with star networks. 

The theoretical predictions were verified by the experiments performed with a complex electrochemical reaction system where multiple bistable elements were connected to a central one.
It should be noted that, although our experimental setup indeed represented a star network of bistable and forced elements, it was not accurately described by the one-component reaction-diffusion model used in the theoretical analysis. 
A realistic quantitative model for such electrochemical elements is not yet available, but it should definitely include many chemical components. 
It is remarkable, though, that this simple theoretical model (without kinetic information about the reaction and with a greatly simplified coupling scheme) is capable of predicting the spatiotemporal dynamics in a complex chemical reaction occurring in star networks. This study thus complements the previously reported findings for the bistable tree networks \cite{Kouvaris:PLOSONE_2012, Kouvaris:ANGIE_2016} and demonstrates a firm understanding of the generic mechanism where a localized perturbation can trigger the formation of stationary patterns in bistable networks.

\begin{acknowledgements}
N.E.K., A.I. and A.D.-G. acknowledge financial support by the MULTIPLEX (Contract No.317532), the MINECO (projects FIS2012-38266 and FIS2015-71582), and the Generalitat de Catalunya (project 2014SGR-608).
N.E.K. also acknowledges support by the HBP SGA1 (Project No.720270).
M.S. and I.Z.K. acknowledge support by the National Science Foundation CHE-1465013.
\end{acknowledgements}

\end{document}